\documentclass[11pt, letterpaper]{article}
\usepackage{amsmath}
\usepackage{graphicx}
\usepackage[a4paper, total={6in, 9.7in}]{geometry}
\usepackage{amsfonts}
\usepackage{booktabs}
\usepackage{array}
\graphicspath{ {./} }
\usepackage{float}
\usepackage{natbib}
\usepackage{xcolor}
\usepackage{hyperref}

\hypersetup{
    colorlinks=true,
    linkcolor=blue,
    citecolor=blue,
    urlcolor=blue
}
\usepackage{lineno}

\begin{document}

\begin{titlepage}

\begin{center}
     \huge{\textbf{A Markov Chain Modeling Approach for Predicting Relative Risks of Spatial Clusters in Public Health}}\\
\vspace{4mm}
\normalsize{\textbf{Lyza Iamrache$^1$, Kamel Rekab$^1$, Majid Bani-Yagoub$^{*1}$,  Julia Pluta,$^2$ and Abdelghani Mehailia$^3$}\\}
\vspace{4mm}
\footnotesize{\textbf{$^1$}Division of Computing, Analytics and Mathematics, School of Science and Engineering, University of Missouri-Kansas City, Kansas City, MO 64110, USA.\\
\textbf{$^2$} School of Law, Cornell University, 340A Myron Taylor Hall, Ithaca, NY 14853-4901,  USA.\\
\textbf{$^3$}Department of Business Administration, Yorkville University, Toronto, ON, M4Y 1W9 Canada}\\

\vspace{2mm}

\footnotesize{\textbf{Emails:} lyzaiamrache@umkc.edu, rekabk@umkc.edu, $^*$ baniyaghoubm@umkc.edu (corresponding author), plutaj@umkc.edu, amehailia@yorkcilleu.ca}
\end{center}
\par\noindent\rule{\textwidth}{0.5pt}
\\

\textbf{Abstract}
Predicting relative risk (RR) of spatial clusters is a complex task in public health that can be achieved through various statistical and machine-learning methods for different time intervals. However, high-resolution longitudinal data is often unavailable  to successfully apply such methods. The goal of the present study is to further develop and test a new methodology proposed in our previous work for accurate sequential RR predictions in the case of limited longitudinal data. In particular, we first use a well-known likelihood ratio test to identify significant spatial clusters over user-defined time intervals. Then we apply a Markov chain modeling  approach to predict RR values for each time interval.  Our findings demonstrate that the proposed approach yields better performance with COVID-19 morbidity data compared to the previous study on mortality data. Additionally, increasing the number of time intervals enhances the accuracy of the proposed Markov chain modeling method. \\

\par\noindent\rule{\textwidth}{0.5pt}
\\

\section{Introduction}
Spatio-temporal modeling has played a critical role in understanding disease transmission dynamics and guiding targeted public health interventions \cite{alqadi2022incorporating, Corkran2025, spatiotemporal_review, Sara2023}. Such models have been applied to detect community-level hot spots, quantify transmission heterogeneity, and assess how demographic and structural factors influence disease burden \cite{spatiotemporal_methods}. These include  spatial--temporal cluster analyses identifying COVID-19 hot spots and their demographic associations \cite{AlQadi2021, AlQadi2023}.  \\

Traditional spatial scan statistics developed by Kulldorff \cite{Kulldorff2006}, were later expanded through spatial scan statistics to identify statistically significant disease clusters after they occur. These methods have been used in many settings, including hospital-acquired infections and avian influenza surveillance. However, because they are retrospective in nature, their ability to predict upcoming waves is limited.  In our previous study \citep{Mortality2025}, we developed a novel Markov chain modeling approach with an embedded corrector to predict the relative risks of spatial clusters using U.S. COVID-19 mortality data across seven distinct time intervals from May 2020 to March 2023. This methodology demonstrated moderate predictive accuracy, offering a promising framework for anticipating areas at heightened risk of severe outcomes during the pandemic. %However, mortality data, while critical for understanding the ultimate impact of a disease, captures only the most severe outcomes and may not fully reflect the early transmission dynamics that are crucial for timely interventions. Morbidity data, which encompasses all confirmed cases of COVID-19, provides a more comprehensive dataset with a significantly larger sample size, potentially offering greater sensitivity to the spatial and temporal patterns of disease spread. This characteristic of morbidity data makes it a more suitable candidate for predictive modeling, as it can capture the initial stages of outbreaks and the broader dissemination of the virus across populations.\\

Building on our previous research, the present study extends the Markov chain modeling framework to predict relative risks of spatial clusters. We use U.S. COVID-19 morbidity longitudinal  data to further examine our proposed approach. In particular, we first use a well-known likelihood ratio test to identify significant spatial clusters over user-defined time intervals .\\

\section{Materials and Methods}

\subsection{Spatio-temporal Data}

Before performing the predictive modeling, we conducted a retrospective spatial analysis of COVID-19 morbidity data in the United States, focusing specifically on confirmed cases.\\
For the analysis, we used a Poisson-based spatial scan model with a cylindrical scanning window. Then we tested the significance of the identified clusters using Monte Carlo Simulation \citep{Besag1991,Block2007,Silva2009}.\\

For the retrospective analysis, we used COVID-19 morbidity data across the United States obtained from the New York Times \cite{NYTimes2021} covering the period May 2020 to March 2023.
The data was split into seven intervals, and each includes his primary variants (alpha, beta, and omicron) that facilitate the time series analysis method. Table 1 presents the split of time intervals, with the primary variants of COVID-19 associated with each sub interval. Note that we chose the  sub-intervals  according to the dominant variants and Mean Case Fatality Rate (CFR). \\

\begin{table}[h!]
\centering
\caption{COVID-19 morbidity data over time interval May 24, 2020 to March 12, 2023, divided into seven sub-intervals  according to the dominant variants and Mean Case Fatality Rate (CFR)}
\label{tab:revised_intervals}
\small % Reduces font size slightly to fit more columns
\begin{tabular}{ccccc}
\hline
\textbf{Interval} & \textbf{Date Range} & \textbf{Primary Variants} & \textbf{Key Sub-lineages} & \textbf{Est. CFR (\%)} \\ \hline
I & 05/24/20 – 09/13/20 & Wild-type, D614G & B.1, B.1.1 & 2.0 -- 4.6 \\
II & 09/13/20 – 03/14/21 & Alpha, Beta & B.1.1.7, B.1.351 & $\sim$2.0 \\
III & 03/14/21 – 06/13/21 & Alpha, Delta & B.1.1.7, B.1.617.2 & 2.5 -- 2.6 \\
IV & 06/13/21 – 10/31/21 & Delta & B.1.617.2, AY.x & $\sim$3.4 \\
V & 10/31/21 – 03/13/22 & Delta, Omicron & BA.1, BA.1.1 & 1.5 -- 1.7 \\
VI & 03/13/22 – 10/16/22 & Omicron & BA.2, BA.4, BA.5 & $<$1.0 \\
VII & 10/16/22 – 03/12/23 & Omicron & BQ.1, XBB.1.5 & $<$0.8 \\ \hline
\end{tabular}
\end{table}

%Intervals I and II, from May 2020 to March 2021, correspond to the emergence of the Alpha and Beta variants. The Delta variant appears in interval III and continues through interval V, covering the period from March 2021 to March 2022. During interval V, the Omicron variant also begins to emerge and remains present until the final interval. \\

%\subsubsection{Advantages of Morbidity Data over Mortality Data}\\

The use of morbidity data is preferred to predict the relative risk over mortality data because the morbidity dataset (i) is larger than mortality data, which allows us to find precise identification of spatial clusters and more reliable estimates of relative risk, and (ii) captures the early transmission, which is better for detecting early disease transmission patterns.

\subsection{Spatial Statistical Scan}
We used SaTScan software to identify spatial clusters, which relies on a Poisson-based spatial scan model \cite{alqadi2022incorporating}. The model employs a cylindrical scanning window, where the circular base defines the geographical scope and the height defines the temporal duration; the maximum cluster size was 25\% of the population at risk \cite{Kulldorff1997, Kulldorff2006}. We used the likelihood ratio test to group clusters with significantly high morbidity risks. given:\\

\begin{equation}
\frac{L(C)}{L_0} = \frac{\left( \frac{n_c}{\mu(c)} \right)^{n_c} \left( \frac{N - n_c}{N - \mu(c)} \right)^{N - n_c}}{\left( \frac{N}{\mu(T)} \right)^N},
\end{equation}

Where $n_c$ is observed cases in cluster $C$, $\mu(c)$ is expected cases in $C$ (under uniform risk), is $N$ = total observed cases, and $\mu(T)$ = total expected cases.\\

We determine the significance of the clusters if their p-value (found using Monte Carlo simulation with 999 iterations) is less than 0.05, where the cluster with the highest likelihood ratio test is considered the primary cluster, and others are secondary clusters based on significance.

\subsection{Markov Chain Modeling Approach}
The prediction of the relative risk in spatial clusters is a complex task that requires a reliable prediction method, which we address using a predictor-corrector approach rooted in Markov chain method \cite{kemeny1976, whittaker1994}, following the methodology used in Estimating the relative risks of spatial clusters using a predictor-corrector method \cite{Mortality2025}. The objective of our method is to find the best prediction of the relative risk of COVID-19 morbidity clusters for the time interval (k+1) using the k previous intervals. \\
%\subsubsection{Model Architecture and Function}\\

We used a prediction approach that consists of selecting a method  between multiple linear regression and exponential smoothing to estimate the corrected future relative risk.  We used inputs $T_1, T_2,..., T_k$ that represent the significant clusters of relative risk in each interval of time to predict $T_{k+1}^*$. Our approach works with all intervals of time compared to Markov chains that use only the most recent relative risk. This methodology is crucial for morbidity data, which provides a better prediction compared to the mortality data.\\

%\subsection{Estimating the Parameters of \( T_{k+1} \)}
To predict, $T_{k+1}^*$ we used a convex combination of the most recent observed risk ($T_k$) and the corrected estimate ($T_k^*$).  which is expressed as:
\begin{equation}
T_{k+1}^* = \alpha^* T_k + (1 - \alpha^*) T_k^*,
\end{equation}
where $0 \leq \alpha^{*} \leq 1$ is the weighting parameter.\\

To find $\mathbf{T_k^*}$ we evaluated the forecasts generated by multiple linear regression ($\hat{T}_k$) and exponential smoothing ($\tilde{T}_k$). We chose the final technique to use in the final model by comparing which technique minimizes the sum of squared errors between the predicted and the observed data. The mean squared error is calculated using this formula:

\begin{equation}
d(T^{*}_{k}, T_{k}) = \sum_{m} \sum_{n} (T^{*}_{k}(m,n) - T_{k}(m,n))^{2},
\end{equation}

Where m represents the row index and n the column index of the matrix.\\

We determined \( \alpha^* \) by minimizing the cumulative sum of squared errors along prior intervals:

\begin{equation}
\sum_{i=3}^{k}{d(\tilde{T}_i(\alpha^*), T_i)} = \min_{0<\alpha<1}{\left\{\sum_{i=3}^{k}{d(\tilde{T}_i(\alpha), T_i)}\right\}}.
\end{equation}
The final formula to predict $T^{*}_{k+1}$ is designed to balance the contribution of recent observations ($T_{k}$) with that of the corrected estimate ($T^{*}_{k}$). In order to include the context of morbidity in the data, we will use the transition matrices and other features, such as population density, healthcare access, and vaccination coverage. \\

%\subsection{Prediction of $T^*_{k+1}$}
The corrected estimate $T^{*}_{k}$ can be obtained through two alternative approaches, each designed to capture patterns in the evolution of transition probabilities across successive intervals. Together, these approaches provide a comprehensive strategy for forecasting $T^{*}_{k+1}$. %3.3.1 \underline{Derivation of \( \tilde{T}_k \) Exponential Smoothing:}
Exponential smoothing is a popular time series forecasting technique based on historical data, where we use the weighted average of past observations.\citep{brown1956}.\\
$\tilde{T}_k$ is computed recursively as follows:
\begin{equation}
\begin{tabular}{l}
$\tilde{T}_k = \alpha T_{k-1} + (1 - \alpha) \tilde{T}_{k-1}$\\
$\tilde{T}_{k-1} = \alpha T_{k-2} + (1 - \alpha) \tilde{T}_{k-2}$\\
$\vdots$\\
$\tilde{T}_3 = \alpha T_2 + (1 - \alpha) \hat{T}_2$
\end{tabular}
\end{equation}

Where, $\hat{T}_2$ is an initial estimate that we obtained using simple linear regression based on $T_1$ and $T_2$. The parameter $\alpha$ determines how much weight we will give to the current observation and the past estimates, with smaller values of $\alpha$ giving more importance to the past predictions.\\

%3.3.2 \underline{Derivation of \( \hat{T}_k \) Multiple Linear Regression:}

Multiple linear regression is a statistical technique used to estimate the relationship between $T_k$ the preceding transition matrices; this approach is used to capture linear trends across successive intervals \citep{burden1985}. The corrected estimate $\hat{T}_k$ is then expressed as:

\begin{equation}
\hat{T}_k = a_0 + \sum_{i=1}^{k-1}{a_i T_i},
\end{equation}
Where, $a_0$ represents the intercept and $a_i$ is the coefficient corresponding to each preceding matrix $T_i$, estimated using the least squares method.\\
We used this approach because it is well suited for morbidity data, as it detects the linear influence of factors such as increasing vaccination coverage or declining case rates over time, which tend to be more pronounced in morbidity data than in mortality data.

\begin{table}[h]
\centering
\caption{Descriptive statistics of identified spatial clusters of COVID-19 morbidity for each time interval I--VII.}
\begin{tabular}{>{\centering\arraybackslash}m{1cm} >{\centering\arraybackslash}m{1.6cm} >{\centering\arraybackslash}m{1cm} >{\centering\arraybackslash}m{2cm} >{\centering\arraybackslash}m{2cm} >{\centering\arraybackslash}m{2cm} >{\centering\arraybackslash}m{3cm}}
\toprule
\textbf{Interval} & \textbf{High Risk} & \textbf{Low Risk} & \textbf{Area of High Risk} & \textbf{Area of Low Risk} & \textbf{Overlap Area} & \textbf{Total Area} \\
\midrule
I & 31 & 42 & 2,268,204.96 (24.80\%) & 4,178,081.51 (45.67\%) & 17,058.77 (0.19\%) & 6,429,228 (70.28\%) \\
II & 39 & 35 & 3,415,409.58 (37.34\%) & 2,495,134.79 (27.28\%) & 18,850.59 (0.21\%) & 5,891,694 (64.41\%) \\
III & 31 & 50 & 1,415,889.82 (15.48\%) & 4,831,282.12 (52.82\%) & 14,880.64 (0.16\%) & 6,232,291 (68.13\%) \\
IV & 38 & 27 & 4,667,511.42 (51.03\%) & 1,406,170.81 (15.37\%) & 9,295.83 (0.10\%) & 6,064,386 (66.30\%) \\
V & 48 & 44 & 2,331,817.55 (25.49\%) & 3,256,131.57 (35.60\%) & 23,603.86 (0.26\%) & 5,564,345 (60.83\%) \\
VI & 42 & 41 & 1,723,199.09 (18.84\%) & 4,306,396.54 (47.08\%) & 23,948.07 (0.26\%) & 6,005,648 (65.65\%) \\
VII & 44 & 44 & 1,559,079.44 (17.04\%) & 3,856,793.51 (42.16\%) & 20,380.57 (0.22\%) & 5,395,492 (58.98\%) \\
\bottomrule
\end{tabular}
\end{table}

\begin{table}[h]
\centering
\caption{Changes of high-risk and low-risk COVID-19 morbidity spatial clusters over user-defined time intervals}
\begin{tabular}{>{\centering\arraybackslash}m{2cm} >{\centering\arraybackslash}m{3cm} >{\centering\arraybackslash}m{3cm} >{\centering\arraybackslash}m{3cm} >{\centering\arraybackslash}m{3cm}}
\toprule
\textbf{Intervals} & \textbf{High-Risk Overlap} & \textbf{Low-Risk Overlap} & \textbf{High-Low Transition} & \textbf{Low-High Transition} \\
\midrule
I $\rightarrow$ II & 768,655.60 33.89\%, 22.51\% & 1,041,575.00 24.93\%, 41.51\% & 433,872.08 19.13\%, 17.39\% & 1,323,277.28 31.67\%, 38.74\% \\
II $\rightarrow$ III & 273,228.68 8.00\%, 19.30\% & 873,705.58 35.02\%, 18.08\% & 1,800,532.85 52.72\%, 37.27\% & 557,223.74 22.33\%, 39.36\% \\
III $\rightarrow$ IV & 446,823.50 31.56\%, 9.57\% & 511,874.79 10.60\%, 36.40\% & 365,132.89 25.79\%, 25.97\% & 2,064,354.90 42.73\%, 44.23\% \\
IV $\rightarrow$ V & 1,251,204.24 26.81\%, 53.66\% & 387,208.61 27.54\%, 11.89\% & 1,356,048.31 29.05\%, 41.65\% & 341,476.65 24.28\%, 14.64\% \\
V $\rightarrow$ VI & 533,307.49 22.87\%, 30.95\% & 1,680,039.56 51.60\%, 39.01\% & 709,727.92 30.44\%, 16.48\% & 278,025.97 8.54\%, 16.13\% \\
VI $\rightarrow$ VII & 391,696.27 22.73\%, 25.11\% & 2,032,047.26 47.19\%, 52.69\% & 303,660.00 17.62\%, 7.87\% & 394,881.91 9.17\%, 25.32\% \\
\bottomrule
\end{tabular}
\end{table}

\section{Results}
%\section{Predictive Modeling of Morbidity Relative Risks}
\subsection{Descriptive Statistics of Spatial Clusters}
This section summarizes the results found from the COVID-19 spatial scan analysis. This analysis is grouped into seven different time intervals to capture the temporal shifts in morbidity patterns influenced by the different variants. Table 2 presents the static cluster counts and the area of high risk (relative risk $> 1$) and the area of low risk (relative risk $< 1$) for each interval. Table 3 presents transition matrices between high-risk and low-risk categories across consecutive time intervals.
%4.1.1 \underline{Exponential Smoothing:}\\
\subsection{Prediction Calculations}
To test the accuracy of our proposed model, we use the relative risk data obtained from the statistical spatial scan, where we want to predict the relative risk of the seventh time interval using the six previous ones. We can check the accuracy of our approach because the seventh interval is known.  Given $T_1, T_2, T_3, T_4, T_5, T_6$ the observations of six previous intervals to find $T^*_7$ the prediction of $T_7$.  We use the exponential smoothing technique to find $\tilde{T}_6$ the predictor of $T_6$ 

\begin{equation}
\begin{tabular}{l}
$\tilde{T}_t = \alpha T_{5} + (1 - \alpha) \tilde{T}_{5}$\\
$\tilde{T}_{5} = \alpha T_{4} + (1 - \alpha) \tilde{T}_{4}$\\
$\vdots$\\
$\tilde{T}_3 = \alpha T_2 + (1 - \alpha) \hat{T}_2$\\

$\hat{T}_2 = 0.457 + 0.556T_1$
\end{tabular}
\end{equation}

$\hat{T}_2$ represents the prediction of $T_2$ using the Multiple linear regression method given $T_1$.\\
$\tilde{T}_6$ was calculated with $\alpha = 0$

\begin{equation}
\tilde{T}_6 = \tilde{T}_5=\tilde{T}_4=\tilde{T}_3=\hat{T}_2,
\end{equation}
The sum of squared errors \( d(\tilde{T}_6, T_6) = 0.6376 \)\\

%4.1.2 \underline{Multiple Linear Regression:}
We use  multiple linear regression to find $\hat{T}_6$ which is also a prediction of ${T}_6$.

\begin{equation}
\hat{T}_6 = 0.0579 + 0.1269T_1 + 0.0424T_2 + 0.0165T_3 + 0.0191T_4 + 0.8206T_5.
\end{equation}
The sum of squared errors between $\hat{T}_6$ and ${T}_6$ $d(\hat{T}_6, T_6) = 0.1183$ is smaller with the sum of squared errors using the exponential smoothing method, indicating a superior fit.\\ 

Since the multiple linear regression method has the superior fit compared to the exponential smoothing method and $\alpha^*=0$, so we can write $T^*_7$ as:

\begin{equation}
T_7^* = \hat{T}_6.
\end{equation}

Figure 1 (a) shows the estimated optimal $\alpha$ by minimizing the cumulative sum of squared errors between $\tilde{T}_k$ and ${T}_k$ across the six intervals $k=1,2,...,6$, the graph shows that $\alpha*=0$. Figure 1 (b) compares our proposed model, multiple linear regression, and the exponential smoothing model. It shows that our model has the best accuracy compared to the other models, since we can see in the graph that it has the greatest coefficient of determination $R^2$. The graph also shows that our model has the smallest MSE.

\begin{figure}[t]
    \centering
    \includegraphics[scale=.45]{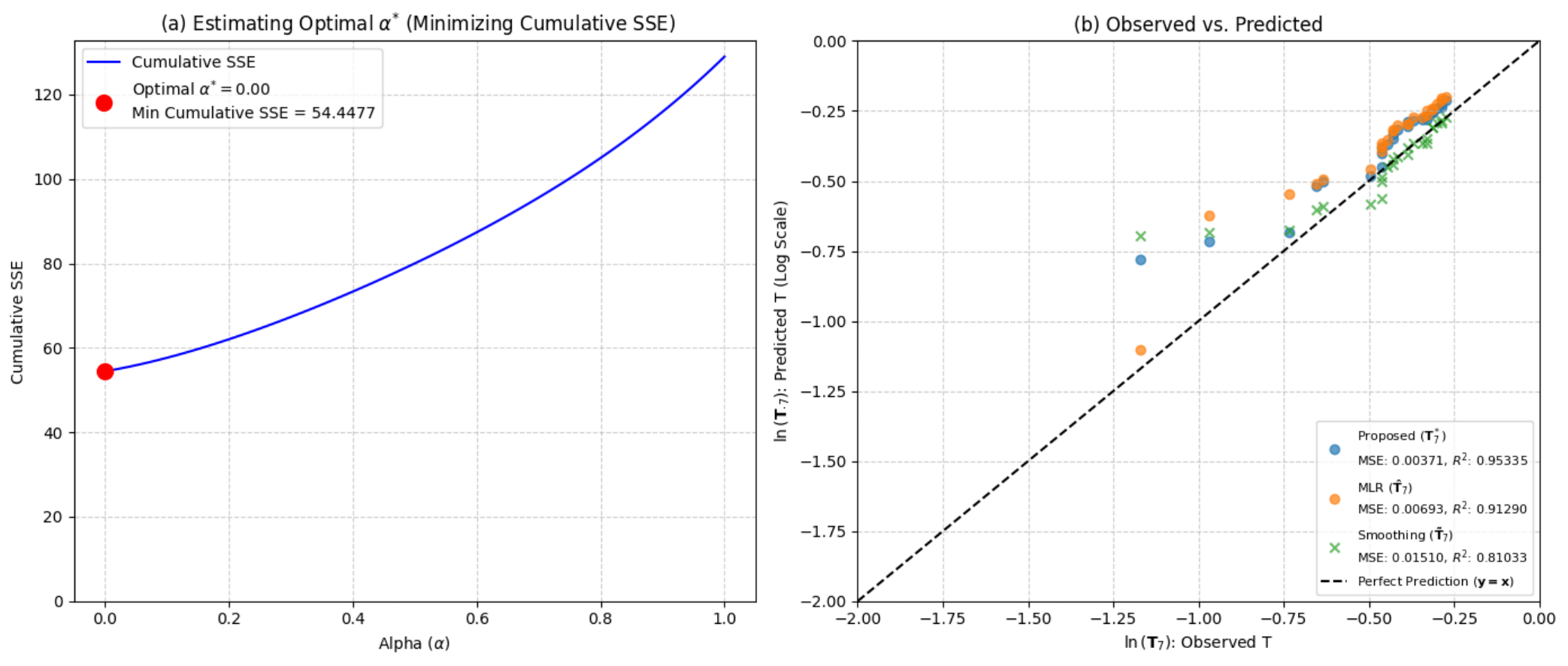}
    \caption{Model validations. (a) Optimal value of $\alpha$  and (b) the accuracy of the proposed model for interval 7 across the multiple regression linear model and the exponential smoothing model. } 
  \label{fig:validation_metrics}
\end{figure}

\subsection{Validation of Predictions}
To check the variability of relative risks across intervals, we calculated the coefficient of variation (CV) for the observed relative risks $T_1$, $T_2$,..., $T_7$ and the predicted relative risks $T^*_7$. The CV formula is given:
\[
\mathrm{CV} = \frac{\mathrm{SD}}{\mathrm{Mean}} \times 100,
\] 
to calculate the relative variability of the relative risks, we provide a standardized metric to compare the consistency of observed and predicted values across intervals. Table 5 reports the mean, standard deviation (SD), and CV for each interval's observed relative risks and the predicted relative risks for interval 7.

 The CV for $T^*_7$ is 29.48\%, lower than that of $T_7$ (32.36\%), indicating that the predicted relative risks  reduce variability compared to the observed values. This reduction in variability suggests that our model, by including additional covariates, smooths out extreme fluctuations in the morbidity data, leading us to more stable and reliable data.
\begin{table}[h!]
\centering\caption{Coefficient of Variation (CV) for observed relative risks \( T_1 \) to \( T_7 \) and predicted relative risks \( T_7^* \), showing mean, standard deviation (SD), and CV for each.}
\begin{tabular}{>{\centering\arraybackslash}m{2cm} >{\centering\arraybackslash}m{2cm} >{\centering\arraybackslash}m{2cm} >{\centering\arraybackslash}m{2cm}}
\toprule
\textbf{Interval} & \textbf{Mean} & \textbf{SD} & \textbf{CV (\%)} \\
\midrule
\( T_1 \) & 0.85 & 0.58 & 68.81 \\
\( T_2 \) & 0.94 & 0.36 & 38.40\\
\( T_3 \) & 0.77 & 0.35 & 45.70 \\
\( T_4 \) & 1.39 & 1.01 & 72.46 \\
\( T_5 \) & 0.90 & 0.20 & 22.53 \\
\( T_6 \) & 0.92 & 0.27 & 29.64 \\
\( T_7 \) & 0.87 & 0.28 & 32.36 \\
\( T_7^* \)  & 0.92 & 0.27 & 29.48 \\
\bottomrule
\end{tabular}
\end{table}
\newpage

\section{Discussion}

The main objective of the present study was to estimate the relative risk of spatial clusters over a sequential time interval. To achieve this, we first identified spatial clusters (e.g. using morbidity longitudinal data by using SaTScan  \cite{Kulldorff1997, Block2007, AlQadi2021}), and then applied the balanced Markov chain approach to refine and predict the relative risk estimates of future time intervals  \cite{Mortality2025}.\\

Our study showed that the balanced Markov chain is highly effective for predicting relative risk of spatial clusters. The model performed predictions close to the observed relative risk, giving a very high $R^2$ value of 0.95682 and a low squared error of 0.00344. In addition to that, the proposed model had a better performance than the exponential smoothing model and multiple linear regression model, where we reached an  $R^2$ of about 0.892 using multiple linear regression and $R^2$ of about 0.810 using the exponential smoothing model. These results show that using a method that combines both multiple linear regression and exponential smoothing is more effective than using either method alone, where exponential smoothing reduces noise and multiple linear regression captures meaningful changes in risk.\citep{burden1985,brown1956}\\

The smoothing parameter $\alpha$ was chosen by minimizing the sum of squared errors, and the optimal value $\alpha^*=0$ indicates  that the model use the past predicted values. The multiple linear regression was used to estimate the previous interval because it had the smallest sum of squared errors compared to the exponential smoothing model that helped to capture short-term and nonlinear changes in morbidity risk, which improved the final prediction.\\

Analyzing the coefficient of variation calculated for both predictive and observed morbidity relative risk, we can conclude that our model produced more accurate risk estimates, as shown in the coefficient of variation table, where the estimated relative risk has a coefficient of variation of 29.48\%, which was lower compared to the observed relative risk with a coefficient of variation of 32.36\%. This result confirmed that the model smooths the extreme variations in the data and provided more consistent risk estimates that can help public health planning and decision-making.\\

Even though our model has a strong performance, like all studies our work also has limitations as follows. First, as mentioned previously, we used SaTScan's Poisson-based spatial scan to identify the spatial clusters, where the approach assumes constant risk within clusters; as result, some spatial information may not be fully captured, which affects the prediction. Second, the Markov chain model is based on the previous time interval to model transitions in relative risk. This is not addressed in the balanced Markov chain, which includes multiple past intervals. This model has similar limitations to Markov-based epidemic models that were mentioned in the previous study \cite{Mortality2025}. Finally, our data was obtained from the New York Times dataset \cite{NYTimes2021} which is publicly accessible. The data can be affected by reporting delays and underreporting, which may influence our results.\\ 

Our work can be extended in the future by adding more detailed covariates or by using adaptive weighting approaches to further improve prediction performance. Overall, our work is based on transforming the spatial scan statistic from a purely retrospective tool into a prediction model by combining a balanced Markov chain approach and morbidity data, which in the future can be extended by using non-Markov dependencies and real-life mobility data to improve prediction performance.

\section*{Data availability}
The data used in this study are publicly available. For the retrospective analysis, we used COVID-19 morbidity data for the United States obtained from The New York Times COVID-19 data repository \cite{NYTimes2021}. These data are openly accessible and can be downloaded from the New York Times GitHub repository.

\bibliographystyle{plain}
\bibliography{refs}

\end{titlepage}

\end{document}